\documentclass[aps,prl,notitlepage,superscriptaddress,twocolumn,showpacs,10pt]{revtex4-1}

\usepackage{natbib}
\usepackage{graphicx}
\usepackage[ansinew]{inputenc}
\usepackage{bm}
\usepackage{color}

\newcommand{\Ktwo}{K$_2$CuF$_4$}

\begin{document}

\title{Ferromagnetic two-dimensional crystals: Single layers of \Ktwo}

\author{B. Sachs}

\affiliation{I. Institut f{\"u}r Theoretische Physik, Universit{\"a}t Hamburg, Jungiusstra{\ss}e 9, D-20355 Hamburg, Germany}

\author{T. O. Wehling}

\affiliation{Institut f{\"u}r Theoretische Physik, Universit{\"a}t Bremen, Otto-Hahn-Allee 1, D-28359 Bremen, Germany}

\affiliation{Bremen Center for Computational Materials Science, Universit{\"a}t Bremen, Am Fallturm 1a, D-28359 Bremen, Germany}

\author{K. S. Novoselov}

\affiliation{School of Physics and Astronomy, University of Manchester, Manchester  M13 9PL, United Kingdom}

\author{A. I. Lichtenstein}

\affiliation{I. Institut f{\"u}r Theoretische Physik, Universit{\"a}t Hamburg, Jungiusstra{\ss}e 9, D-20355 Hamburg, Germany}

\author{M. I. Katsnelson}

\affiliation{Radboud University of Nijmegen, Institute for Molecules and Materials, Heijendaalseweg 135, 6525 AJ Nijmegen, The Netherlands}

\begin{abstract}

The successful isolation of graphene ten years ago has evoked a rapidly growing scientific interest in the nature of two-dimensional (2D) crystals. A number of different 2D crystals has been produced since then, with properties ranging from superconductivity to insulating behavior. Here, we predict the possibility for realizing ferromagnetic 2D crystals by exfoliating atomically thin films of \Ktwo. From a first-principles theoretical analysis, we find that single layers of \Ktwo\ form exactly 2D Kosterlitz-Thouless systems. The 2D crystal can form a free-standing membrane, and exhibits an experimentally accessible transition temperature and robust magnetic moments of 1$\mu_B$ per formula unit. 2D \Ktwo\ unites ferromagnetic and insulating properties and is a demonstration of principles for nanoelectronics such as novel 2D-based heterostructures.

\end{abstract}

\maketitle

\normalsize

\textit{Introduction} The large variety of two-dimensional (2D) atomic crystals which are available to us \cite{novoselov2005two,novoselov2011nobel} comprise into a very rich class of materials, which collectively covers a large diversity of properties. Also, the recent progress in creating heterostructures made from individual atomic crystals \cite{dean2010boron,ponomarenko2011tunable,haigh2012cross} allowed the investigation of such phenomena as Coulomb drag \cite{gorbachev2012strong} and the Hofstadter butterfly \cite{ponomarenko2013cloning,dean2013hofstadter,hunt2013massive}, but also facilitated the creation of novel, often multifunctional devices, such as tunneling transistors \cite{britnell2012field,georgiou2012vertical} and photovoltaic devices \cite{britnell2013strong,sachs2013doping,zhang2013ultrahigh}.

One property which has been missing, however, is the ferromagnetism, which remains an elusive phenomenon in 2D crystals for a number of reasons, including chemical instabilities of such compounds and the nature of the Kosterlitz-Thouless transition. It would be extremely interesting to obtain stable 2D ferromagnetic crystals both for the purposes of experiments on Kosterlitz-Thouless transitions and for possible applications in heterostructures (such as spin-polarized electrodes for spintronic applications in heterostructures based on 2D atomic crystals).

Here, we propose to realize ferromagnetic 2D crystals by exfoliating monolayers of \Ktwo. Based on our \textit{ab initio} theoretical studies, we show that the structural properties of 2D \Ktwo\ are preserved under exfoliation and find cleavage energies comparable to other layered materials that allow for exfoliation, such as graphite. We demonstrate the Kosterlitz-Thouless type of ferromagnetism in 2D \Ktwo\ and calculate the transition temperature. 2D \Ktwo\  does not require substrate support, turns out to be robust in the magnetism and is therefore highly promising for novel applications as well as fundamental studies of the Kosterlitz-Thouless transition in a real 2D material.

\begin{figure*}[htb]
\includegraphics[width=0.99\textwidth]{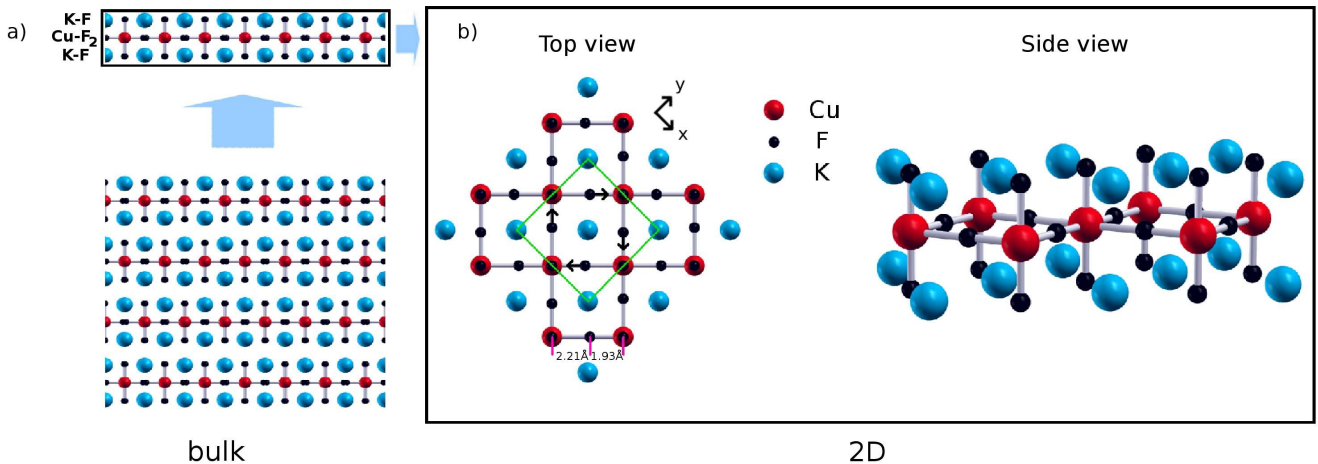}
\caption{(a) Crystal structure of bulk \Ktwo\ from a side view. (b) Crystal structure of 2D \Ktwo\ from the top view and side view. The green lines mark the unit cell, and the distances between the purple lines show different Cu-F bond lengths due to Jahn-Teller distortions (black arrows), which do not differ by more than 1 pm in the bulk and the monolayer. For clarity, only Cu-F bonds are visualized by gray lines.}
\label{fig:struct}
\end{figure*}

\Ktwo\ is a layered double perovskite that belongs to the class of K$_2M$F$_4$ ($M$=Mn, Fe, Co, Ni, Cu) \cite{eyert1993electronic} materials. Its crystal structure was an object of intense research more than three decades ago \cite{Hirakawa1973, Hidaka1983343, khomskii1973orbital}. The bulk system is built up from layers with a respective thickness of three atoms in an alternate stacking (Fig. \ref{fig:struct}a). Thereby, one layer is composed of a Cu-F$_2$ plane sandwiched by two K-F planes. Each Cu atom is thus surrounded by six F atoms, four in the plane, and two above and below the Cu atom in the K-F planes. Due to Jahn-Teller distortions of the F atoms, the tetragonal symmetry of the bulk crystal is lifted such that in the end an orthorhombic lattice symmetry is present \cite{Hidaka1983343, khomskii1973orbital}.

\textit{Cleavage, structure, and stability.} We first discuss why an exfoliation of 2D layers of \Ktwo\ is feasible and how this affects the atomic structure and the stability of the isolated layer. Under exfoliation of 2D \Ktwo\ from the bulk crystal, so-called cleavage decohesion energies have to be overcome which are determined by the strength of the interlayer binding. Before calculating the properties of cleavage fracture, we tested explicitly the quality of our density functional theory simulations for the intact bulk system. We were able to reproduce the experimentally known atomic positions in the bulk system including Jahn-Teller distortions of 2.5\% of the lattice constant (for details, see Ref. \footnote[1]{For additional information on computational details, an extended discussion of the orbital-related magnetism, and elastic property calculations, supplementary information material is provided.}).  

\begin{figure}[htb]
\includegraphics[width=0.99\columnwidth]{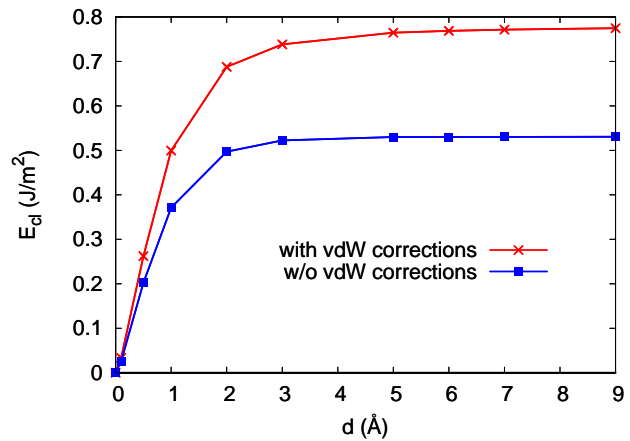}
\caption{Cleavage energy $E_{\rm cl}\left(d\right)$ as a function of the separation $d$ of the two crystal halves. $d$ is defined with respect to the layer-layer separation of 2.6 \AA\ in the equilibrium configuration.}
\label{fig:cleavage}
\end{figure}

In order to simulate the exfoliation procedure, we introduced a fracture in the bulk.  We then calculated total energies under variation of the separation $d$ between the fractured parts. The resulting cleavage energy $E_{\rm cl}\left(d\right)$ is shown in Fig. \ref{fig:cleavage}. The ideal cleavage cohesion energy \cite{medvedeva1996first} is obtained from the asymptotic limit of $E_{\rm cl}\left(d\right)$. We find values of 0.78 J/m$^2$ (including van der Waals corrections) and 0.53 J/m$^2$ (without van der Waals corrections). This is about 1.5 - 2 times higher than the experimentally estimated cleavage energy in graphite \cite{zacharia2004interlayer}. The maximum derivative of  $E_{\rm cl}\left(d\right)$ gives the so-called theoretical cleavage strength \cite{medvedeva1996first}. Here we find 5.7 GPa (with vdW) and 4.4 GPa (w/o vdW). 

We also calculated binding energies of the isolated bilayer system. We obtain binding energies virtually the same as the ideal cleavage energies, indicating that the interaction between \Ktwo\ and a substrate does not depend on the number of \Ktwo\ layers. Compared to other 2D crystals, the bilayer binding is about two to three times larger than in graphite \cite{lebegue2010cohesive, liu2012interlayer} and 1.5 - 2 times larger than in Bi$_2$Se$_3$ \cite{seixas2013topological}, as obtained from theory. Both materials \cite{novoselov2004electric,hong2010ultrathin} and many other \cite{novoselov2005two,bonaccorso2012production} allow for the isolation of 2D crystals by mechanical exfoliation techniques, and we infer from our calculations that the interlayer binding in \Ktwo\ is weak enough to exfoliate single layers of \Ktwo\ as well. Similar as for other 2D crystals, a possible way to produce single layers of \Ktwo\ could be the Scotch tape method.

Afterwards, we checked for possible a structural modification of the isolated single layer which is highly relevant for the magnetism of the system. Here, we found the atomic positions of K and Cu atoms in the isolated layer to be virtually the same as in the bulk system and also the Jahn-Teller distortions of F atoms were restored, but with slightly different Cu-F bond lengths than in the bulk (Fig. \ref{fig:struct}b). With these distortions, atoms in single-layer \Ktwo\ are arranged in a quasi-2D square unit cell consisting of two formula units (four K, two Cu, and eight F atoms, see the green box in Fig. \ref{fig:struct}b).

\begin{figure}[htb]
\includegraphics[width=0.99\columnwidth]{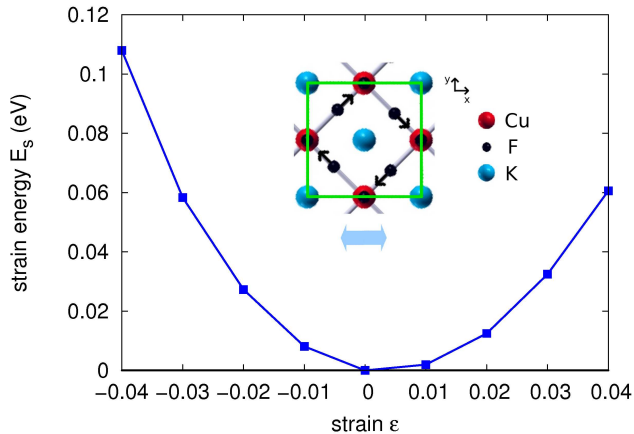}
\caption{Elastic energy of \Ktwo\ under axial strain. (Inset) Visualization of axial strain. The square unit cell (green box) is stretched or compressed in the x direction.}
\label{fig:strain}
\end{figure}

A crucial question for future experiments will be whether 2D sheets of \Ktwo\ are suitable to form free-standing membranes. In an exfoliation procedure, large flakes of arbitrary shape are formed, and a high in-plane stiffness is needed to avoid curling and to allow the material to withstand its own weight or external loads.
In order to judge the in-plane stiffness of \Ktwo, we performed simulations of axial strain effects.  The strain energy curve is shown in Fig. \ref{fig:strain}. One can see a strong tension-compression asymmetry. From the strain energy curves we determine the in-plane stiffness by the 2D Young's modulus which we define as
\begin{equation}
Y_{\rm 2D}=\frac{1}{A_0} \frac{\partial^2 E_{\rm s}}{\partial \epsilon^2} \Big|_{\epsilon=0},
\end{equation} 
with $\epsilon$ the axial strain, $E_{\rm s}$ the total strain energy (per unit cell), and $A_0$ the equilibrium surface. We find $Y_{\rm 2D}=44.8$ N/m, which is about 13\% of the in-plane stiffness in the ultrastrong material graphene (see Ref. \cite{booth2008macroscopic}).  From this rather high in-plane stiffness, we can deduce from elastic theory calculations (see Ref. \footnotemark[1]) that even for large flakes, 2D \Ktwo\ is able to withstand its own weight, significantly large extra loads, or vibration during  handling, demonstrating that the membrane properties are sufficient to produce a free-standing \Ktwo\ membrane without the support of a substrate. 

It is important to note that the Jahn-Teller distortions are robust under in-plane strains. Their presence is decisive for the magnetism, which we investigate below.\\

\textit{Magnetism and Kosterlitz-Thouless transition.} We now discuss the peculiarities of the magnetism of 2D \Ktwo. In the past decades, a lot of effort has been put into the research of magnetic materials, including finite size scaling of ferromagnetism in thin films (see, e.g., Refs. \cite{farle1987ferromagnetic,huang1993finite,waldfried1998finite}). These materials require a supporting substrate to ensure structural stability.

Single layers of \Ktwo, however, form a stable, truly 2D system and represent a different case: Due to the Mermin-Wagner theorem, long-range ordering at finite temperature is expected to be forbidden for an exactly 2D system with spin-rotational symmetry. On the other hand, the Kosterlitz-Thouless transition, which is a quasi-long-range ordering effect in 2D systems, could make experiments on a ferromagnetic 2D crystal feasible. Indeed, in bulk \Ktwo, the existence of a critical temperature $T_{\rm C}$ as well as the Kosterlitz-Thouless transition have been observed.  In order to realize ferromagnetism experimentally also in the 2D crystal, three conditions need to be fulfilled: 

1. At zero temperature, ferromagnetic coupling of the bulk \cite{khomskii1973orbital,eyert1993electronic} must be preserved in the 2D layer.  2. The magnetism should be robust against deformations and strains.  3. The Kosterlitz-Thouless temperature must be sufficiently high. 

\begin{figure}[htb]
\includegraphics[width=0.99\columnwidth]{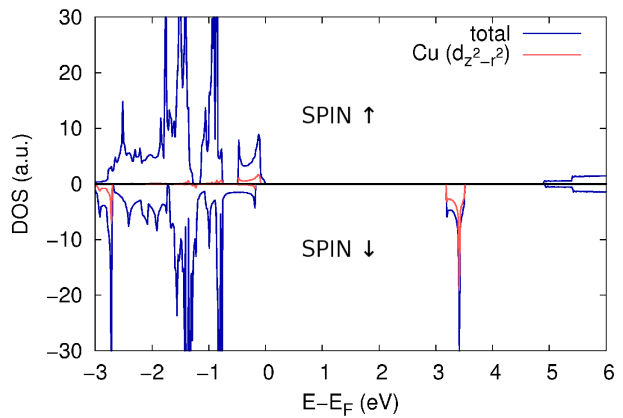}
\caption{Total DOS of the system (blue). The pronounced hole resonance is mostly of a Cu $d_{\rm z^2-r^2}$ character (red).}
\label{fig:DOS}
\end{figure}

The first two conditions are closely related to the robustness of Jahn-Teller distortions. We find from our simulations that the fullly relaxed system is ferromagnetic due to the formation of stable magnetic moments of 1$\mu_{\rm B}$ per formula unit. The Jahn-Teller distortions (and therefore the magnetic moments) also survived the application of axial strains. This stability of magnetic moments is remarkable and has not been found in a 2D crystal so far. Only in artificially strained NbSe$_2$ and NbS$_2$ has a formation of magnetic moments been predicted theoretically \cite{NbS2strain}. The reason for the stability of the ferromagnetism in 2D \Ktwo\ at low temperature can be understood best from the electronic structure. The spin-polarized density of states (DOS) obtained from the local density approximation with the Hubbard $U$ correction is shown in Fig. \ref{fig:DOS}a. A large band gap can be observed in which we find located a pronounced hole state at 3.2 eV in the spin-down channel. Similar as in the bulk case, the hole state in the spin-down channel instead is mostly of a $d_{\rm z^2-r^2}$ character (with admixtures of hybridized Cu $d_{\rm xy}$ and F $p$ states) and is responsible for the ferromagnetic ordering by an orbital ordering effect \cite{farle1987ferromagnetic},\footnotemark[1]. This hole state turns out to be robust under strains as well. For the considered axial strains, it changes position by some hundred meV, but not enough to become quenched.

The third condition of a finite transition temperature is intimately connected with the magnetic anisotropy. It is known from the experimental literature that bulk \Ktwo\ exhibits a small magnetic in-plane anisotropy \cite{yamazaki1981plane}: The easy direction of the magnetization axis lies in-plane but without any preferred direction within the plane. Although the magnetic interlayer coupling in the bulk is known to be weak (less than 0.1\% of the intralayer coupling \cite{Hirakawa1973}), a chemical interaction between the layers is present and it is not clear how the anisotropy evolves in a monolayer. Therefore, we performed first-principles calculations of the magneto-crystalline anisotropy energy (MCAE) calculations in 2D \Ktwo\ (see Ref. \footnotemark[1]) and found the same easy axis as in the bulk and even stronger anisotropy (see below). Hence, single-layer \Ktwo\ is a 2D spin-1/2 ferromagnet with the spin moments arranged on a square lattice in an alignment parallel to the $x$-$y$ plane. To our best knowledge, this is the first realization of a truly 2D Kosterlitz-Thouless magnetic system.

From a model point of view, this system can be described by an effective Heisenberg model for which we determine in the following exchange parameters from our first-principles calculations. In analogy to Ref. \cite{irkhin1999kosterlitz}, we describe the magnetism of the system by an effective easy-plane Heisenberg Hamiltonian
\begin{equation}
H=-\frac{1}{2} \sum_{\langle i,j \rangle} J \left[ S_i^x S_j^x +  S_i^y S_j^y + \eta S_i^z S_j^z \right],
\end{equation}
with $J>0$ being the nearest-neighbor ferromagnetic exchange and $1-\eta \ll1$ describing a small in-plane anisotropy.  From density functional theory (DFT) total energy calculations we derive model parameters $J$ and $\eta$: $J=E_{\rm{fm}}-E_{\rm{afm}}$ with $E_{\rm{fm}}$ the total energy of the ferromagnetic system and $E_{\rm{afm}}$ the total energy of the (virtual) antiferromagnetically ordered solution. $\eta$ is directly related with the MCAE. Using the LDA+$U$ method \cite{LichtensteinLDAU}, we obtain the parameters $J/k_{\rm B}=25.3$ K and $\eta=0.90$, which we use to calculate the Kosterlitz-Thouless temperature ($T_{\rm KT}$).  A formula for $T_{\rm KT}$ has been derived from a renormalization group analysis in Ref. \cite{irkhin1999kosterlitz}:
\begin{equation}
T_{\rm KT}=\frac{2 \pi J S^2}{\ln\left(\sqrt{\frac{T_{\rm KT}/J S}{1-\eta}}\right)+2 \ln\left(\frac{2}{T_{\rm KT}/2 \pi J S^2}\right)+C},
\end{equation}
with $C=-0.5$ (from bulk). This finally yields $T_{\rm KT} = 7.9$K, a value comparable to the critical temperature in the bulk ($T_{\rm C, bulk}\sim 6.3$K).  Hence, all conditions to realize ferromagnetism in a free-standing layer of \Ktwo\ are fulfilled, which allows for experiments on an exactly 2D Kosterlitz-Thouless system.

\textit{Conclusion \& outlook.} We have demonstrated the possibility to produce ferromagnetic 2D crystals from \Ktwo. 2D \Ktwo\ exhibits a stable ferromagnetic ground state below 8 K, a sizable band gap, and the stiffness to form free-standing membranes of a large area, where confinement effects are marginal. This allows to perform experiments on a truly 2D Kosterlitz-Thouless system without the perturbance of 3D interactions. For instance, it would be extremely interesting to perform systematic studies of doping effects (e.g., through substituent atoms) in order to increase the Kosterlitz-Thouless temperature. As a 2D crystal with both insulating and ferromagnetic properties, \Ktwo\ opens perspectives for a number of applications, e.g.,  as a magnetic spacer in novel 2D heterostructures. 2D \Ktwo\ can be interfaced with graphene to induce magnetic moments that can be tuned by an external field, or combined with topological insulators in order to create a gauge field acting on Dirac fermions \cite{katsnelson2013plane} for the purpose of band gap opening.

\textit{Acknowledgments.}
Support from the DFG (Germany) via Priority Programme 1459 ``Graphene" and the European Graphene Flagship are acknowledged. \\

\end{document}

\title{Supplementary information for \\ Ferromagnetic two-dimensional crystals: Single layers of \Ktwo}

\author{B. Sachs}

\affiliation{I. Institut f{\"u}r Theoretische Physik, Universit{\"a}t Hamburg, Jungiusstra{\ss}e 9, D-20355 Hamburg, Germany}

\author{T. O. Wehling}

\affiliation{Institut f{\"u}r Theoretische Physik, Universit{\"a}t Bremen, Otto-Hahn-Allee 1, D-28359 Bremen, Germany}

\affiliation{Bremen Center for Computational Materials Science, Universit{\"a}t Bremen, Am Fallturm 1a, D-28359 Bremen, Germany}

\author{K. S. Novoselov}

\affiliation{School of Physics and Astronomy, University of Manchester, Manchester M13 9PL, United Kingdom}

\author{A. I. Lichtenstein}

\affiliation{I. Institut f{\"u}r Theoretische Physik, Universit{\"a}t Hamburg, Jungiusstra{\ss}e 9, D-20355 Hamburg, Germany}

\author{M. I. Katsnelson}

\affiliation{Radboud University of Nijmegen, Institute for Molecules and Materials, Heijendaalseweg 135, 6525 AJ Nijmegen, The Netherlands}

%\pacs{72.80.Rj; 73.20.Hb; 73.61.Wp}

%\thanks{}

%\date{\today}

\maketitle

\section{Computational details}
First-principles density functional theory calculations were carried out using the Vienna \textit{ab initio} simulations package (VASP) \cite{kresse1994norm, kresse1999ultrasoft} within the projector-augmented wave method \cite{Bloechl_PAW}. 

\textbf{Atomic structure.} 
For calculations of the atomic structure of bulk and 2D \Ktwo\, the local density approximation (LDA) was employed to the exchange-correlation potential \cite{KohnShamLDA} because we found that this perfectly reproduces the experimentally known Jahn-Teller distortions. A k-mesh of 8x8x1 (5x5x5) grid points was used for single layer (bulk) \Ktwo\ and a kinetic energy cutoff of 700 eV (single layer) and 500 eV (bulk). We used the experimentally known unit cell volume of the bulk system and the same lattice parameters for the 2D system (lattice constants $a=b=5.8655$ \AA\ \cite{Hidaka1983343}). The atomic positions were relaxed until forces were smaller than 0.02 eV/\AA.

\textbf{Cleavage energies, elastic theory.}
The binding between \Ktwo\ layers is partially due to van der Waals interaction. LDA is known to reproduce the interlayer distances in weakly bound layered systems while GGA fails which we also see in our calculations. Therefore, we performed calculations using LDA. We also employed the LDA with van der Waals corrections within the DFT-D2 method of Grimme \cite{grimme2006semiempirical}. Local Coulomb interactions do not play a role for the binding of \Ktwo\ layers which we tested using the LDA+$U$ method \cite{LichtensteinLDAU}.

\begin{figure*}[htb]
\includegraphics[width=0.3\textwidth]{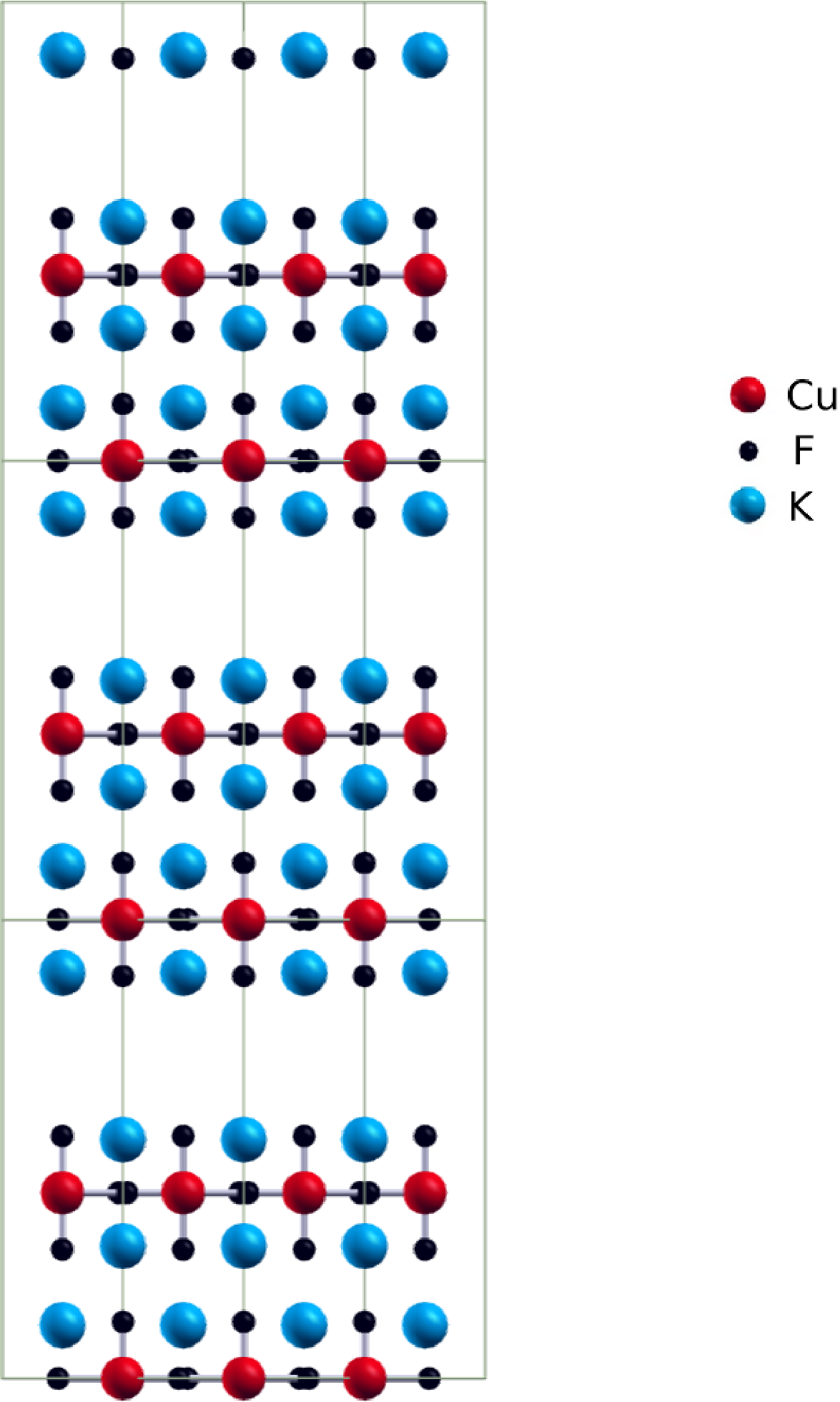}
\caption{Scheme of the geometry used to simulated cleavage fracture. Unit cells are marked by grey lines.}
\label{fig:cleavagestruct}
\end{figure*}

For calculations of the bilayer binding energy, we simulated a bilayer system and relaxed the structure to the equilibrium configuration. Afterwards, we lifted one layer far to the vacuum. The binding energy was then calculated by the difference of total energies in the equilibrium position and at far interlayer distance.

For cleavage energy calculations, we used a bulk unit cell as shown in Fig. \ref{fig:cleavagestruct}. One cell contained two \Ktwo\ layers, which ensured a distance between two fractures of more than 10 \AA\ and the fracture was simulated by increasing the c parameter of the cell. 

For calculations of the elastic energies, LDA+$U$ calculations were performed with the optimized lattice constant $a=b=5.561$ \AA. This lattice constant differs by about 5\% from the experimental value but also reproduces Jahn-Teller distortions and ferromagnetism.

\textbf{Magnetism and electronic structure.}
For electronic structure calculations and calculations of magnetic moments, we account for local Coulomb interactions in Cu $d$ orbitals by making use of the LDA+$U$ method. This also reproduced best the experimentally known magnetism and magnetic coupling of the bulk. In the simple rotationally invariant formulation of Dudarev \cite{Dudarev}, an effective Hubbard parameter $U_{\rm eff}= U_{\rm H} - J_{\rm H}$ is defined, thus only the difference between the average Hubbard Coulomb repulsion $U_{\rm H}$ and Hund's rule exchange $J_{\rm H}$ is important. We chose $U_{\rm eff}=7.03$ eV, a similar value was used to calculate KCuF$_3$ \cite{LichtensteinLDAU}. We emphasize here that the band gap strongly depends on $U_{\rm eff}$, for LDA without Coulomb interactions we find only 0.3 eV instead of 3.2 eV using LDA+U with $U_{\rm eff}=7.03$ eV. For calculations of the magnetic crystalline anisotropy, we included spin-orbit coupling in the calculations and used a denser k-mesh (18x18x1). With the parameter $U_{\rm eff}$, we avoid the problem of a strong anisotropy dependence of the chosen Hund's exchange $J_{\rm H}$ \cite{bousquet2010j}. The anisotropy parameter might slightly differ in reality, but note that the magnetic anisotropy $\eta$ only enters logarithmically in the formula for $T_{\rm KT}$. We also stress here that the shape anisotropy does not significantly contribute to the overall magnetic anisotropy: while the MCAE energy is on the order of $10^{-4}$ eV, we estimate shape anisotropy contributions to be much smaller ($\sim 10^{-6}$ eV). 

\section{Band Structure and orbital ordering}

\begin{figure}[htb]
\includegraphics[width=0.99\textwidth]{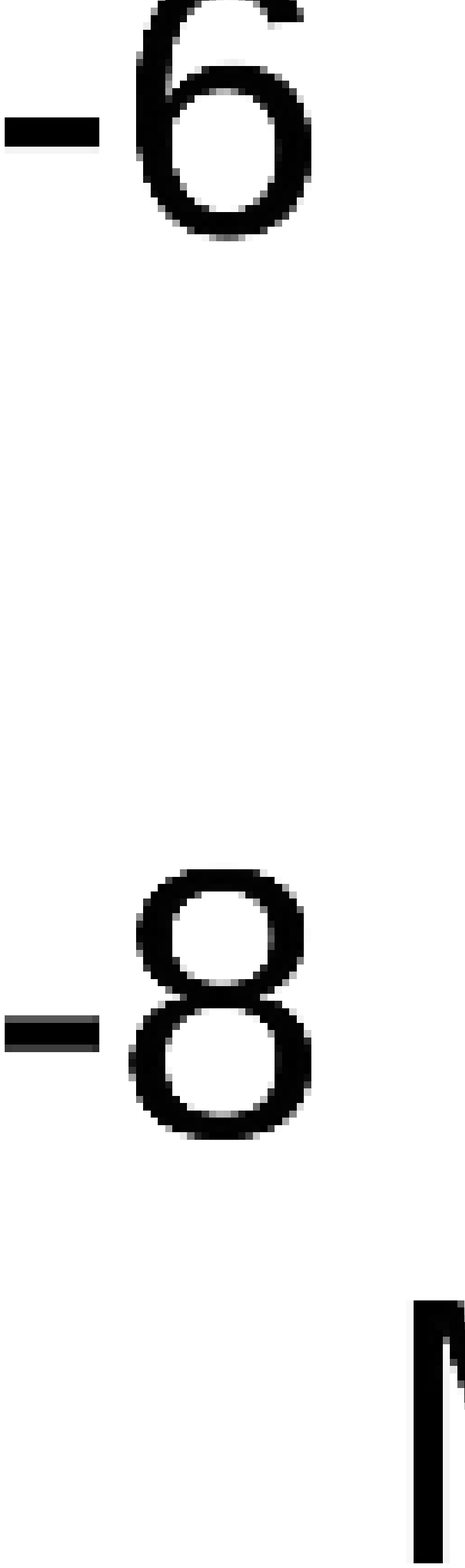}
\caption{a), b) Band structure of single layer \Ktwo\ in the spin up (a) and spin down (b) channel. Red "fat bands" depict the $d_{\rm{z^2-r^2}}$ contribution to the bands, whereby the thickness is proportional to the magnitude of the contribution. c) Contour plot of the hole density in the Cu-F$_2$ plane of the hole state at $\sim 3.5$ eV shown in b) in a window between 0.35--0.55e/unit cell. Red dots denote Cu atoms. }
\label{fig:bands}
\end{figure}

In the manuscript, we state that the magnetism mainly originates from $d_{\rm z^2-r^2}$ orbitals. To learn more about this, we look at the band structure of \Ktwo\ which is depicted in Fig. \ref{fig:bands}a) and b). A comparison of the up and down channels shows that the band gap is direct. Many bands are present below the Fermi level because of the relatively large unit cell, but here we focus on Cu $d_{\rm{z^2-r^2}}$ states. The red fat bands visualize the Cu $d_{\rm{z^2-r^2}}$ character of the bands whereby the thickness of the red band depicts the magnitude of the contribution. One can see a different distribution of spectral weight in both channels. Most importantly, the bands in the spin up channel exhibit a pronounced $d_{\rm{z^2-r^2}}$ contribution at bands below -6 eV that is not present in the down channel. Here, on the contrary, the main spectral weight is located at the hole state.

This hole state is responsible for a ferromagnetic ordering through the following mechanism: the Jahn-Teller distortions of F atoms lead to an orbital ordering effect at Cu sites with an alternating occupation of $d_{\rm{z^2-x^2}}$ and  $d_{\rm{z^2-y^2}}$ orbitals of neighbored Cu atoms. Thus, the ferromagnetism in 2D \Ktwo\ is orbital-selective and related with the $d_{\rm{z^2-x^2}}$ and  $d_{\rm{z^2-y^2}}$ subspace. The orbital ordering is visualized in Fig. \ref{fig:bands}c. We tested explicitly the importance of Jahn-Teller distortions. Indeed, in a fictitious undistorted geometry, we find from our calculations that an antiferromagnetic coupling of spins is energetically slightly more favorable than a ferromagnetic coupling. Only the distortions of F atoms lead to an alternating occupation of the $d_{\rm{z^2-x^2}}$ and  $d_{\rm{z^2-y^2}}$ orbitals which finally allows for a ferromagnetic ordering. This is also in line with a scenario proposed  by Khomskii and Kugel \cite{khomskii1973orbital} to explain ferromagnetic ordering in bulk \Ktwo.

\section{Membrane properties}

%In order to judge the in-plane stiffness of \Ktwo, we calculated elastic energies for uniaxial strains. For this purpose, the 2D unit cell was stretched and contracted along one axis K-F !!!
%
%\begin{figure}[htb]
%\includegraphics[width=0.5\textwidth]{elastic.eps}
%\caption{Elastic energy of \Ktwo\ under uniaxial strain.}
%\label{fig:strain}
%\end{figure}

In Ref. \cite{booth2008macroscopic}, the peculiar elastic properties of graphene were discussed. The authors showed that these allow for fabrication of long graphene beams which need support from only one side due to their extraordinary in-plane stiffness. In the spirit of Ref. \cite{booth2008macroscopic}, we estimate the bending of a \Ktwo\ flake using elastic theory for large flakes of similar length $l$ and width $w$. The typical out-of-plane deformation $h$ of a flake can be estimated by the formula $h/l \approx \left(\rho g l / Y_{\rm 2D}\right)^{1/3}$, with $l$ being the length of the flake, and $\rho=2.1 \times 10^{-6} kg/m^2$ the density. For a large flake of length of the order $l \approx 100$ $\mu$m, we obtain 
$h/l \approx 10^{-3}-10^{-4}$. For graphene, $h/l \approx 10^{-4}$ for flakes of this size. Thus, even for large flakes, 2D \Ktwo\ is able to withstand its own weight, significantly large extra loads or vibration during handling.